# Terramechanics-Inspired Sensor Fusion for Slip and Mass Flow Estimation in Agricultural Conveyor Systems

**Ruben Hefele**, **Timo Oksanen**

*Technical University of Munich, Germany; Chair of Agrimechatronics; Munich Institute of Robotics and Machine Intelligence (MIRMI) (e-mail: ruben.hefele@tum.de).*

**Abstract:** Precision farming requires accurate knowledge of implement states. In organic fertilizer spreaders, heterogeneous material and variable field conditions complicate flow-rate prediction, and tilt angles alter the slip behaviour between transport floor and material. This paper presents a model-based sensor fusion method for slip and mass-flow estimation, integrating weighing, conveyor-speed, and inertial measurements in an Extended Kalman Filter with a volumetric flow model and a longitudinal dynamics model. Validation on real-world cattle-manure spreading shows mass estimates tracking the raw weight with an RMSE of 11.9 kg and integrated mass flow within 8.4 % of the observed mass loss.

## 1. INTRODUCTION

Precision farming in modern agricultural operations requires accurate application and documentation of products such as seeds and fertilizers. Mineral fertilizers are comparatively easy to manage because they are industrially manufactured and therefore exhibit well-defined, homogeneous properties that have been characterised in test halls. In contrast, the application of organic fertilizers is more complex due to large heterogeneity in material properties – both between farms and even within a single trailer load. For solid manure in particular, its characteristics depend on for example livestock type, straw usage, moisture and the storage time of the material, all of which influence its consistency and flow behaviour (Malgeryd and Wetterberg, 1996).

Solid manure is typically applied using trailers equipped with a hopper, a scraper conveyor for unloading (Landry et al., 2005), and a rear-mounted spreading unit designed to shred and distribute the material as evenly as possible across the working width (EN 13080:2002). Controlling the mass flow through the spreading unit – such as achieving tractor speed independent constant or variable application rates – requires modulation of conveyor speed, commonly driven by a hydraulic motor (Landry et al., 2005; Balsari et al., 2015). Slip between the material and conveyor slats complicates accurate mass flow prediction (Kosch et al., 2004; Landry et al., 2006). Although modern high-end spreaders include weighing units for real-time hopper monitoring (Deter, 2016), estimating the real-time slip behaviour remains a challenge.

Landry (2005) measured mass flow on a stationary manure spreader using four load cells between hopper and frame, deriving the rate in post-processing from 2 Hz absolute-mass recordings filtered by a fifth-order low-pass filter; the study targeted manure–machine interaction modelling rather than real-time measurement. Miclet et al. (2010) proposed a two-stage control strategy: the discharge-gate position and scraper-conveyor speed are jointly controlled by two inductive sensors that detect the force exerted by the product on the gate, while axle and drawbar strain gauges provide 10 s weight estimates used for low-frequency density calibration rather than closed-loop flow control. Both approaches achieve acceptable constant application rate but lack the real-time estimation of internal states – instantaneous mass flow and material–floor slip – required for dynamic variable-rate control.

Building on findings that beater drive power is proportional both to conveyor power (Thirion et al., 1998) and to mass flow (Hügle, 1999), Kosch et al. (2004) used beater torque as feedback to modulate conveyor speed, with material-specific torque curves removing the need for a weighing unit. However, validation against collection-tray measurements showed no significant flow-rate improvement, as torque remained nearly constant except near empty hopper states where decreasing bulk density made the torque signal useful for application-rate determination. In orchard applications, Balsari et al. (2015) estimated mass flow from pre-calibrated material transport per chain-conveyor motor revolution, achieving high accuracy and rapid response with separated pig-slurry solids, though slip was not accounted for and the authors

⋆ The project is supported by funds of the Federal Ministry of Food and Agriculture (BMEL) based on a decision of the Parliament of the Federal Republic of Germany. The Federal Office for Agriculture and Food (BLE) provides coordinating support for digitalisation in agriculture as funding organisation, grant number 28DE112B18.

noted that other manure types, such as cattle manure may behave differently.

Material properties such as cohesion and internal friction angle are critical for modelling solid-manure flow behaviour and can be quantified by direct shear testing (Landry, 2005; Thirion et al., 1998; Malgeryd and Wetterberg, 1996; Landry et al., 2005).

Slip modelling in longitudinal off-road vehicle dynamics is well established in terramechanics (Renius, 2019), where tractive force arises from plastic soil deformation under longitudinal shear. For tracked vehicles, the Bekker–Wong formulation models this force as a function of normal stress, slip, cohesion, and internal friction angle (Wong, 2022).

For mining-sector conveyors, Rehman et al. (2017) applied a linear Kalman filter to noisy tachometer data and fused the estimated belt velocity with a moving-average-smoothed load-cell signal to obtain mass flow at 10 Hz, without accounting for slip.

This paper develops and validates a real-time estimator for material–conveyor slip and mass flow during the unloading process of a chain-conveyor manure spreader. The approach integrates a volumetric flow model into an Extended Kalman Filter (EKF) that fuses measurements of load mass, conveyor speed, and a slip signal derived from a longitudinal dynamics model. This model considers frictional resistance along the walls and floor, the beater-induced loading force, and a tractive drive force. Importantly, the drive force – transmitted from conveyor to load – is modelled using the Bekker–Wong terramechanics formulation (Wong, 2022), adapted to conveyor–material interactions. The methodology is validated using log data collected under real-world operating conditions with cattle manure.

## 2. MATERIALS

A Bergmann universal spreader was used in this study (Fig. 1). It is equipped with a chain conveyor, two horizontal beaters, and a V-spread centrifugal disc spreading unit. The spreader is compatible with ISOBUS (ISO 11783) for variable rate and section control (Ludwig Bergmann GmbH, 2025). Key specifications relevant to the methods employed in this study are summarised in Tab. 1.

While the beaters and spreading discs are powered by the tractor power take-off, the chain conveyor and discharge gate are driven hydraulically via the tractor power-beyond system. Both use closed-loop control, with tachometer feedback for conveyor speed and position feedback for gate clearance. The gate was held at a fixed clearance of 1.4 m throughout this study.

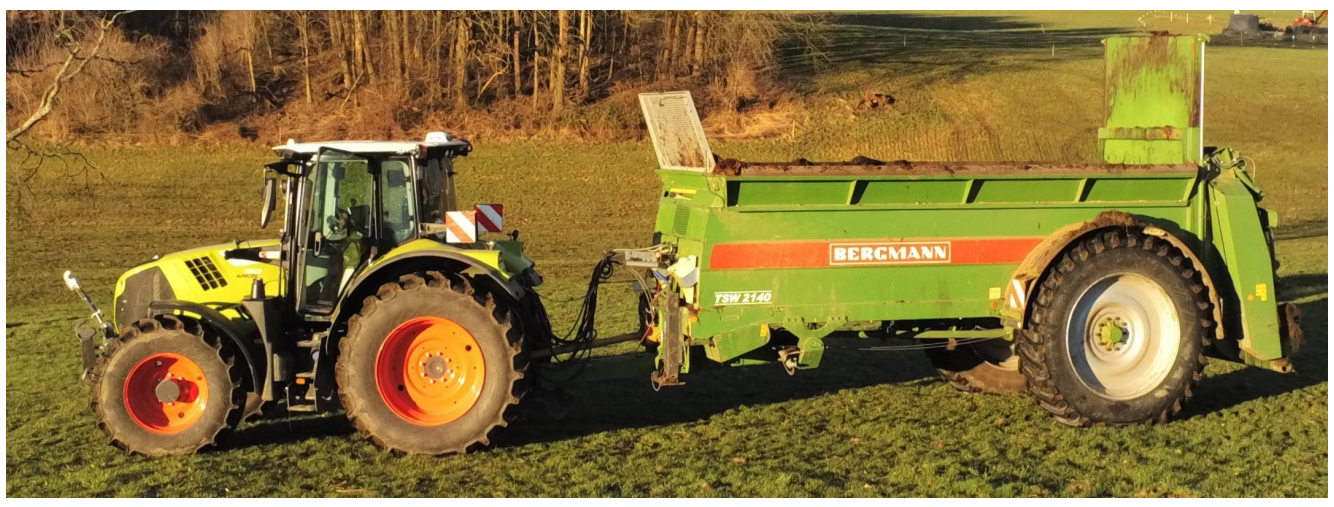

Fig. 1. Claas Arion 550 CMATIC towing the Bergmann TSW 2140E universal spreader on grassland.

Table 1. Spreader parameters used in the slip and flow models.

| Description | Value | Unit |
|---|---|---|
| Length of the hopper | 5.23 | m |
| Width of the hopper | 1.80 | m |
| Height of the hopper | 1.32 | m |
| Max. discharge gate clearance height | 1.40 | m |
| Number of slats (on top of floor) | 16 | – |
| Width of slats | 50 | mm |
| Length of slats | 850 | mm |
| Width of chain | 40 | mm |
| Number of chains | 4 | – |

The spreader features a weighing unit based on load cells. Raw weight data are transmitted via CAN bus at 10 Hz, while conveyor speed data are available at 40 Hz. Additionally, a Gemac Motus XB inertial measurement unit (IMU) provides internally Kalman-filtered (GEMAC Chemnitz GmbH, 2022) measurements of the roll and pitch Euler angles at 20 Hz.

The experiments were conducted using solid manure collected from a deep-litter barn housing breeding cows for approximately 150 days. The bedding was supplemented daily with cut straw at a rate of $1\,\mathrm{kg \cdot m^{-2}}$. Before field application, the manure was relocated, stored for two weeks, and then loaded. Analysis of the material indicated a dry matter content of 34.19 % and a bulk density of $750\,\mathrm{kg \cdot m^{-3}}$.

## 3. METHODS

The estimation procedure is illustrated in Fig. 2. Sensor measurements of mass and floor speed, together with a derived slip signal, are fused within an EKF to estimate the load speed and mass flow. To prevent an algebraic loop, the slip signal is passed through a unit delay. The EKF incorporates a volumetric flow model, described in detail in Sec. 3.1. To obtain the slip signal, a longitudinal dynamics model of the material in the hopper is employed, accounting for the tractive interaction between the transport floor and the load. The slip model is discussed in Sec. 3.2, and the complete EKF implementation is presented in Sec. 3.3.

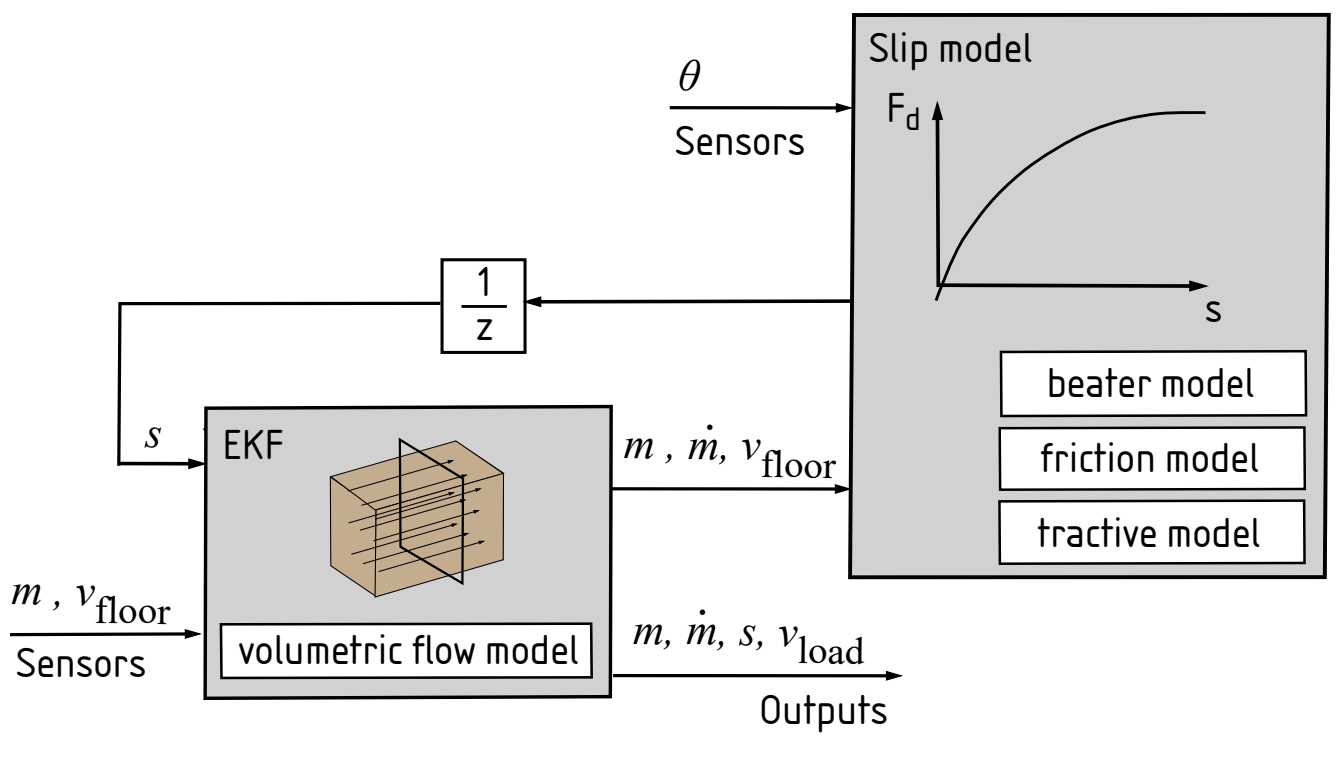

Fig. 2. Estimator architecture. Inputs are mass $m$, floor speed $v_{\mathrm{floor}}$, and IMU-derived pitch $\theta$; outputs include smoothed $m$, mass flow $\dot{m}$, slip $s$, and load speed $v_{\mathrm{load}}$.

### 3.1 *Volumetric flow model*

The volumetric model in Eq. 1 relates mass change to flow through an opening, assuming constant bulk density $\rho$, such that the mass flow rate $\dot{m}$ follows directly from the volumetric flow. At the outlet, material of velocity $v_{\text{load}}$ passes through a rectangular cross-section bounded by the gate clearance $h_{\text{gate}}$ and hopper width $w_{\text{hopper}}$:

$$\dot{m} = \rho\, h_{\text{gate}}\, w_{\text{hopper}}\, v_{\text{load}} \tag{1}$$

The model treats the material as a uniform continuum, ignoring granular fluctuations, and assumes the gate region is always filled, keeping the effective cross-section constant.

### 3.2 *Longitudinal dynamics model as a base for slip estimation*

To initiate and sustain the motion of the load toward the beaters, the driving force must exceed the sum of the resisting forces. Since this force is dependent on slip, the limiting factor is typically not the torque of the hydraulic motor, but the maximum transferable driving force between transport floor and load. The relationship between these forces is quantified using a longitudinal dynamics model of the load (Fig. 3), formulated based on Newton's second law (Eq. 2). Translational dynamics are treated as a point mass, while the load-floor contact is modelled as a rectangular area with uniformly distributed pressure.

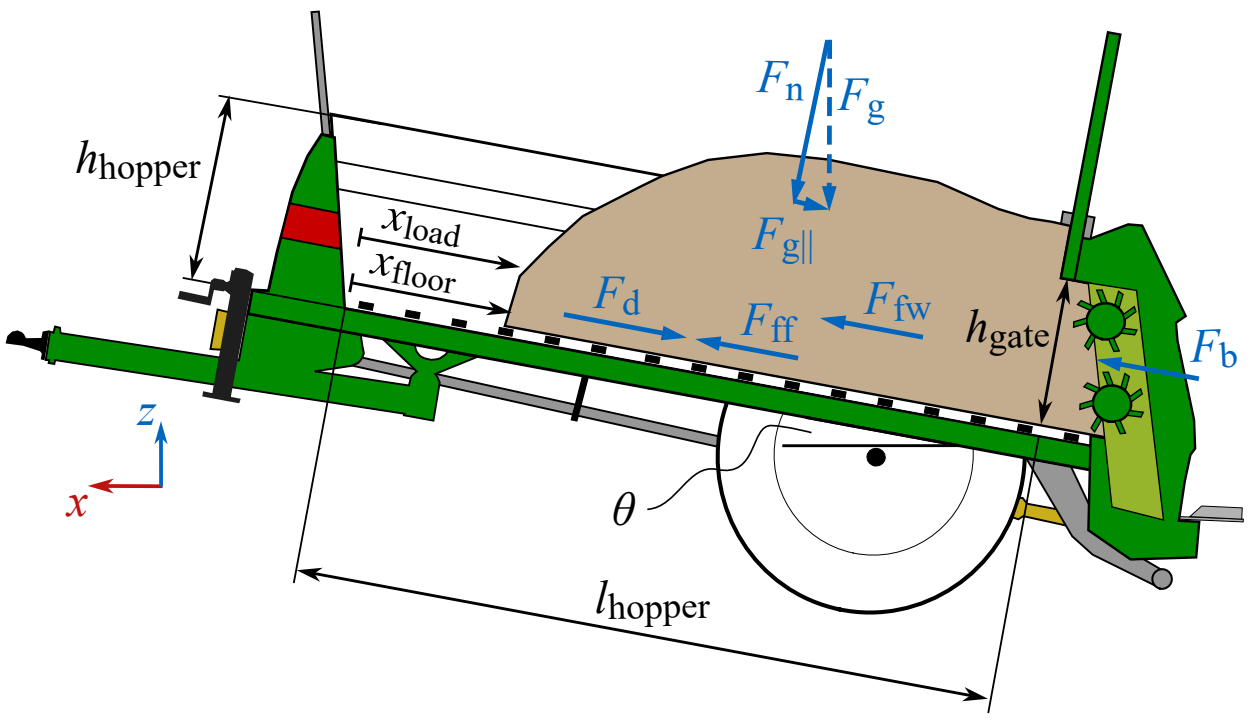

Fig. 3. Longitudinal dynamics model with driving force $F_{\text{d}}$, downslope force $F_{\text{g}\|}$, floor friction $F_{\text{ff}}$, wall friction $F_{\text{fw}}$, beater resistance $F_{\text{b}}$, normal force $F_{\text{n}}$, and gravity $F_{\text{g}}$. Shown further are the pitch angle $\theta$, gate clearance $h_{\text{gate}}$, hopper length $l_{\text{hopper}}$ and height $h_{\text{hopper}}$. Floor and load positions $x_{\text{floor}}$ and $x_{\text{load}}$ are measured from the hopper front.

The longitudinal dynamics model is formulated via force equilibrium (Fig. 3), with load mass $m$, gravitational acceleration $g$, spreader pitch $\theta$, and load acceleration $a_{\text{load}}$:

$$F_{\text{d}} = F_{\text{ff}} + 2\, F_{\text{fw}} + F_{\text{b}} + m\, a_{\text{load}} - F_{\text{g}\|} \tag{2}$$

The downslope force is

$$F_{\text{g}\|} = m\, g\, sin(\theta). \tag{3}$$

By applying Coulomb friction, the floor resistance friction

$$F_{\text{ff}} \leq \mu_{\text{dyn}}\, m\, g\, \cos(\theta)\,(1-f), \quad \text{with } \mu_{\text{dyn}} = k_1\, \mu_{\text{stat}} \tag{4}$$

is obtained. Here, $k_1$ is a tuning parameter introduced to reflect the generally observed reduction of dynamic friction relative to the static friction $\mu_{\text{stat}}$ (Tab. 2). Only the moving parts of the floor (slats and chains, Fig. 4) contribute to friction, captured by the area ratio

$$f = \frac{A_{\text{effective}}}{A_{\text{floor}}} = \frac{2\, n_{\text{slats}}\, w_{\text{slat}}\, l_{\text{slat}} + n_{\text{chains}}\, w_{\text{chain}}\, l_{\text{hopper}}}{w_{\text{hopper}}\, l_{\text{hopper}}}. \tag{5}$$

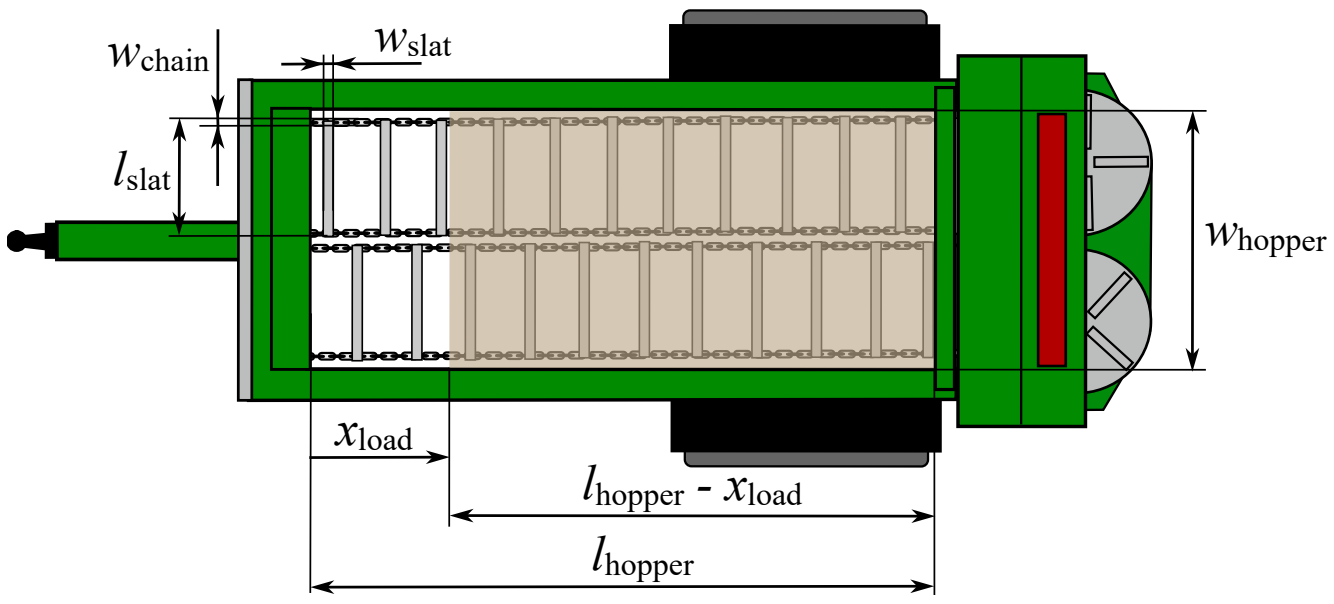

Fig. 4. Spreader top view with chain conveyor. The load-floor contact area is shown in brown.

Table 2. Solid manure characteristics (Landry et al., 2005)

| Static Friction Coefficient on Steel | Angle of Internal Friction | Apparent Cohesion |
|---|---|---|
| [-] | [°] | [kPa] |
| $\mu_{stat}$ | $\phi$ | $c$ |
| 0.73 | 19.3 | 23.0 |

Product characteristics used in the slip model, taken from Landry et al. (2005) and considered representative of the material in this study based on similarities in description, density, and dry matter content.

Coulomb friction also forms the basis of the model for the wall friction force

$$F_{\text{fw}} \leq \mu_{\text{dyn}}\,(l_{\text{hopper}} - x_{\text{load}})\, h_{\text{hopper}}\, p_{\text{mean}}, \tag{6}$$

which uses the same dynamic friction coefficient $\mu_{\text{dyn}}$ as Eq. 4. The load position $x_{\text{load}}$ is approximated from the current mass relative to the initial mass:

$$x_{\text{load}} = l_{\text{hopper}} \left(1 - \frac{m}{m_0}\right). \tag{7}$$

The wall friction acts over the effective wall area (Fig. 5), with mean lateral pressure obtained from the height-

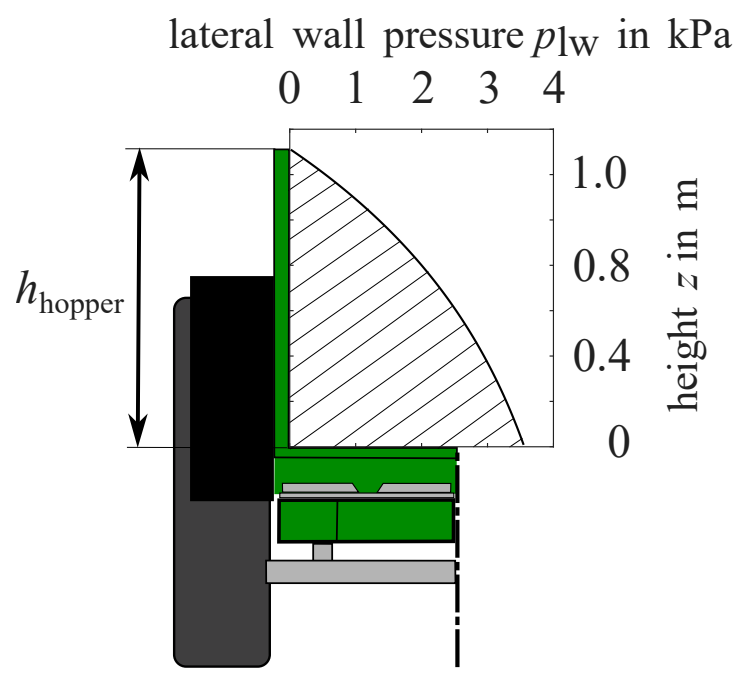

Fig. 5. Lateral wall pressure (rear view) of the spreader, $z$ origin at hopper bottom. Flat, symmetric walls with identical pressure on both sides are assumed.

averaged Janssen distribution (Holický et al., 2009; DIN EN 1991-4:2024):

$$p_{\text{lw}}(z) = \rho\, g\, \frac{R_{\text{h}}}{\mu_{\text{stat}}} \left(1 - \exp\left(-\frac{z\, K\, \mu_{\text{stat}}}{R_{\text{h}}}\right)\right), \tag{8}$$

$$p_{\text{mean}} = \frac{1}{h_{\text{hopper}}} \int_0^{h_{\text{hopper}}} p_{\text{lw}}(z)\, dz, \tag{9}$$

with $R_{\text{h}} = w_{\text{hopper}}\, h_{\text{hopper}}/(w_{\text{hopper}} + 2\, h_{\text{hopper}})$ and $K = 1.1\,(1-\sin\phi)$. Dynamic effects on the lateral pressure, such as those induced by roll motion, are neglected.

The beater force, $F_{\text{b}}$, is assumed to be proportional to the mass flow rate, $\dot{m}$, through the beaters:

$$F_{\text{b}} = k_2\, \dot{m}, \tag{10}$$

where $k_2$ is a proportionality constant.

The transferable force between the load and the transport floor is modelled as a function of slip using the Bekker–Wong formulation (Eq. 11), as established in terramechanics for track–soil interactions (Renius, 2019). This formulation predicts zero transferable force at zero slip, while the force saturates at the maximum value.

$$F_{\text{d}} = (A_{\text{effective}}\; c + F_{\text{n}}\; \tan\phi) \left(1 - \exp\left(-\frac{j}{k_3}\right)\right) \tag{11}$$

Herein, cohesion $c$, internal friction angle $\phi$, and $k_3$, a parameter accounting for the initial slope, are employed. Additionally, the effective contact area used for force transmission is defined as:

$$A_{\text{effective}} = f\; (l_{\text{hopper}} - x_{\text{load}})\; w_{\text{hopper}}, \tag{12}$$

where $f$ (Eq. 5) is the area ratio between the moving parts and the stationary part of the floor. Eq. 13 defines the shear displacement as the product of the effective contact length and slip, $s$ (Baek et al., 2022).

$$j = s\, (l_{\text{hopper}} - x_{\text{load}}) \tag{13}$$

The normal force follows as

$$F_{\text{n}} = m\, g\, \cos(\theta). \tag{14}$$

To sustain motion at acceleration $a_{\text{load}}$, the driving force must satisfy:

$$F_{\text{d}} \geq F_{\text{ff}} + 2\, F_{\text{fw}} + F_{\text{b}} + m\, a_{\text{load}} - F_{\text{g}\parallel}. \tag{15}$$

Applying a nominal acceleration $a_{\text{load}} = 0.01\,\text{m}\cdot\text{s}^{-2}$, slip is obtained by resolving the Bekker–Wong model for $s$:

$$s = \frac{-k_3}{l_{\text{hopper}} - x_{\text{load}}} \ln\left(1 - \frac{F_{\text{d}}}{A_{\text{effective}}\, c + F_n \tan\phi}\right), \tag{16}$$

subject to $F_d \leq (A_{\text{effective}}\, c + F_n \tan\phi)$.

The tuning parameters $k_1$, $k_2$, and $k_3$ were calibrated on replayed field data, yielding $k_1 = 0.7$ (dynamic-to-static friction ratio), $k_2 = 109$ (beater-force proportionality), and $k_3 = 0.25$ (Bekker–Wong slip slope).

### *3.3 Extended Kalman filter*

An EKF is implemented to fuse measurements of mass, $m$, and floor speed, $v_{\text{floor}}$, in order to estimate the mass flow, $\dot{m}$, while providing smoothed values for slip, $s$, and load speed, $v_{\text{load}}$. Following the standard EKF formulation (Bar-Shalom et al., 2001), state $\boldsymbol{x}_k$ is propagated through $\boldsymbol{f}$ with process noise $\boldsymbol{w}_k$ and corrected via observation model $\boldsymbol{h}$ with measurement noise $\boldsymbol{\nu}_k$; no control input is applied ($\boldsymbol{u} = \boldsymbol{0}$). The algorithm operates in discrete time with a sample period of 100 ms, corresponding to the 10 Hz update rate of the weight sensor, which provides the slowest measurement.

The volumetric model (Sec. 3.1) defines the system dynamics within the EKF with the internal states

$$\mathbf{x} = [m\;\; v_{\text{load}}\;\; s]^T\,, \tag{17}$$

while discrepancies between the model and the actual system are corrected using the measurements

$$\mathbf{z} = [m\;\; v_{\text{floor}}\;\; \hat{s}]^T\,, \tag{18}$$

provided by the load cells, the transport floor tachometer, and as a derived measurement $\hat{s}$ by the slip model (Sec. 3.2). This approach preserves the system model in the prediction step, while the measurement updates correct the state estimates, resulting in smoothed signals with minimal delay.

Treating $v_{\text{load}}$ and $s$ as quasi-stationary, the state transition model reads

$$\frac{d\boldsymbol{x}}{dt} = \boldsymbol{f}(\boldsymbol{x}, \boldsymbol{u}) = \begin{pmatrix} \rho\, h_{\text{gate}}\, w_{\text{hopper}}\, x_2 \\ 0 \\ 0 \end{pmatrix}. \tag{19}$$

The measurement model is given as

$$\hat{\boldsymbol{z}} = \begin{pmatrix} m \\ v_{\text{floor}} \\ s \end{pmatrix} = \begin{pmatrix} m \\ \dfrac{v_{\text{load}}}{1-s} \\ s \end{pmatrix} = \boldsymbol{h}(\boldsymbol{x}, \boldsymbol{u}) = \begin{pmatrix} x_1 \\ \dfrac{x_2}{1-x_3} \\ x_3 \end{pmatrix}, \tag{20}$$

where the slip, $s$, is defined as

$$s = \frac{v_{\text{floor}} - v_{\text{load}}}{v_{\text{floor}}}. \tag{21}$$

The EKF is configured with the following process and measurement noise covariance matrices, $\mathbf{Q}$ and $\mathbf{R}$:

$$\mathbf{Q} = \begin{bmatrix} 0.04 & 0 & 0 \\ 0 & 6.4\times10^{-9} & 0 \\ 0 & 0 & 2.5\times10^{-5} \end{bmatrix}, \quad \mathbf{R} = \begin{bmatrix} 9 & 0 & 0 \\ 0 & 1\times10^{-10} & 0 \\ 0 & 0 & 4\times10^{-4} \end{bmatrix} \tag{22}$$

The $\mathbf{R}$ entries for the weighing unit and transport-floor tachometer were set from observed noise variances, the small tachometer value reflecting its high resolution. The remaining $\mathbf{R}$ entry (for the derived measurement $\hat{s}$) and the process noise covariance $\mathbf{Q}$ were tuned on replayed field data to balance the slip model against the sensor measurements.

## 4. RESULTS

Data collected during real-world operation were processed with the proposed algorithm. The transport-floor control strategy enforced a fixed ratio between transport-floor speed and tractor ground speed using a field-calibrated constant slip value and a working width of 25 m. Logged measurements were replayed time-synchronously so the estimator operated chronologically with access only to past data.

A start-up sequence to steady-state floor speed on nearly level terrain is shown in Fig. 6, starting at roughly 40 % fill level. Slip saturates at 0.5. The estimated mass flow $\dot{m}_{\text{est}}$ remains nearly constant and agrees after 4 s with the noisier $\dot{m}_{\text{calc}}$, which is derived from differences of the weight signal over 1 s intervals and therefore delayed.

The force decomposition shows floor friction $F_{\text{ff}}$ dominating, followed by wall friction $F_{\text{fw}}$; beater $F_{\text{b}}$ and downslope $F_{\text{g}\parallel}$ contributions are negligible on level terrain. The mass estimate $m_{\text{est}}$ is effectively smoothed compared to the raw

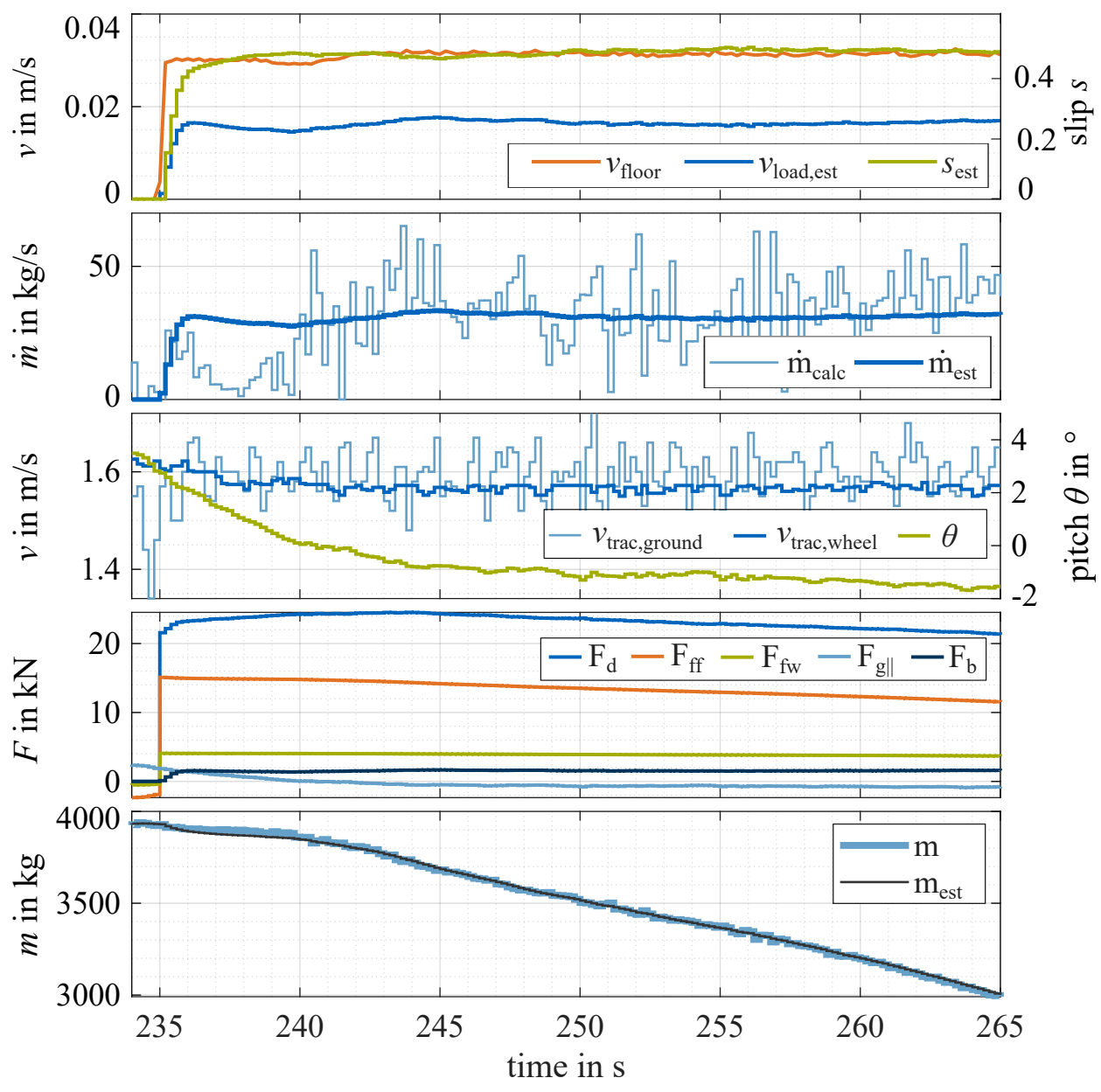

Fig. 6. Start-up to steady-state floor speed on nearly level terrain; target application rate $10\,\mathrm{t}\cdot\mathrm{ha}^{-1}$.

Table 3. Mass-tracking performance of the EKF against the raw weight signal, for three scenarios shown in Figs. 6, 7, and 8.

| Scenario | RMSE [kg] | bias [kg] | $\Delta_{\mathrm{int}}/\lvert\Delta m\rvert$ [%] |
|---|---|---|---|
| Level start-up (Fig. 6) | 11.9 | −0.1 | +1.9 |
| Downhill (Fig. 7) | 20.9 | +12.7 | +22.3 |
| Full load cycle (Fig. 8) | 16.4 | −3.5 | −8.4 |

RMSE and bias compare $m_{\mathrm{est}}$ with $m_{\mathrm{raw}}$. $\Delta_{\mathrm{int}}/\lvert\Delta m\rvert$ is the relative difference between the integrated estimated mass flow, $\int \dot{m}_{\mathrm{est}}\,dt$, and the observed total mass loss, $\lvert m_{\mathrm{raw}}(t_{\mathrm{end}}) - m_{\mathrm{raw}}(t_{\mathrm{start}})\rvert$, over the scenario.

value $m$ while still accurately following its overall trend, with an RMSE of 11.9 kg and an integral consistency within 1.9 % of the observed mass loss (Tab. 3).

A downhill start-up sequence is shown in Fig. 7, where $m_{\mathrm{est}}$ again tracks the raw measurement with an RMSE of 20.9 kg (Tab. 3). On slopes steeper than $-8°$, the downslope force $F_{\mathrm{g}\parallel}$ surpasses the wall friction $F_{\mathrm{fw}}$, and slip increases from 0.3 to 0.6 starting at a pitch of $-6°$ as the load is pushed away from the beaters. Consistent with this, the estimated mass flow $\dot{m}_{\mathrm{est}}$ decreases despite an increase in tractor speed at 35 s. The integral consistency of 22.3 % over the segment, noticeably larger than on level terrain, suggests that the slip-model parameters, calibrated primarily on level-terrain data, may require refinement for steep-slope operation.

A complete load cycle (Fig. 8) represents realistic A–B guidance-line operation with two automatic-section-control headland turns. For most of the time, a nearly constant tractor speed was maintained whenever engine power allowed, with speed increasing as the load emptied after 275 s and the beaters required less power. The driving force $F_d$, along with floor friction $F_{\mathrm{ff}}$ and wall friction $F_{\mathrm{fw}}$, decrease over the operation. Meanwhile, the estimated mass flow $\dot{m}_{\mathrm{est}}$ remains mostly constant within a margin of $10\,\mathrm{kg}\cdot\mathrm{s}^{-1}$. Fluctuation in the estimated slip $s_{\mathrm{est}}$ is caused by an increase in downslope force around 50 s. Overall, slip increases as the fill state decreases. The smoothed mass estimate $m_{\mathrm{est}}$ closely follows the raw measurement $m$ with an RMSE of 16.4 kg and an integral consistency within 8.4 % over the complete cycle (Tab. 3). At near-empty fill states, however, the estimator performs less accurately, as material begins to crumble and no longer covers the full gate cross-section, which is reflected in the estimated mass flow $\dot{m}_{\mathrm{est}}$ after 315 s.

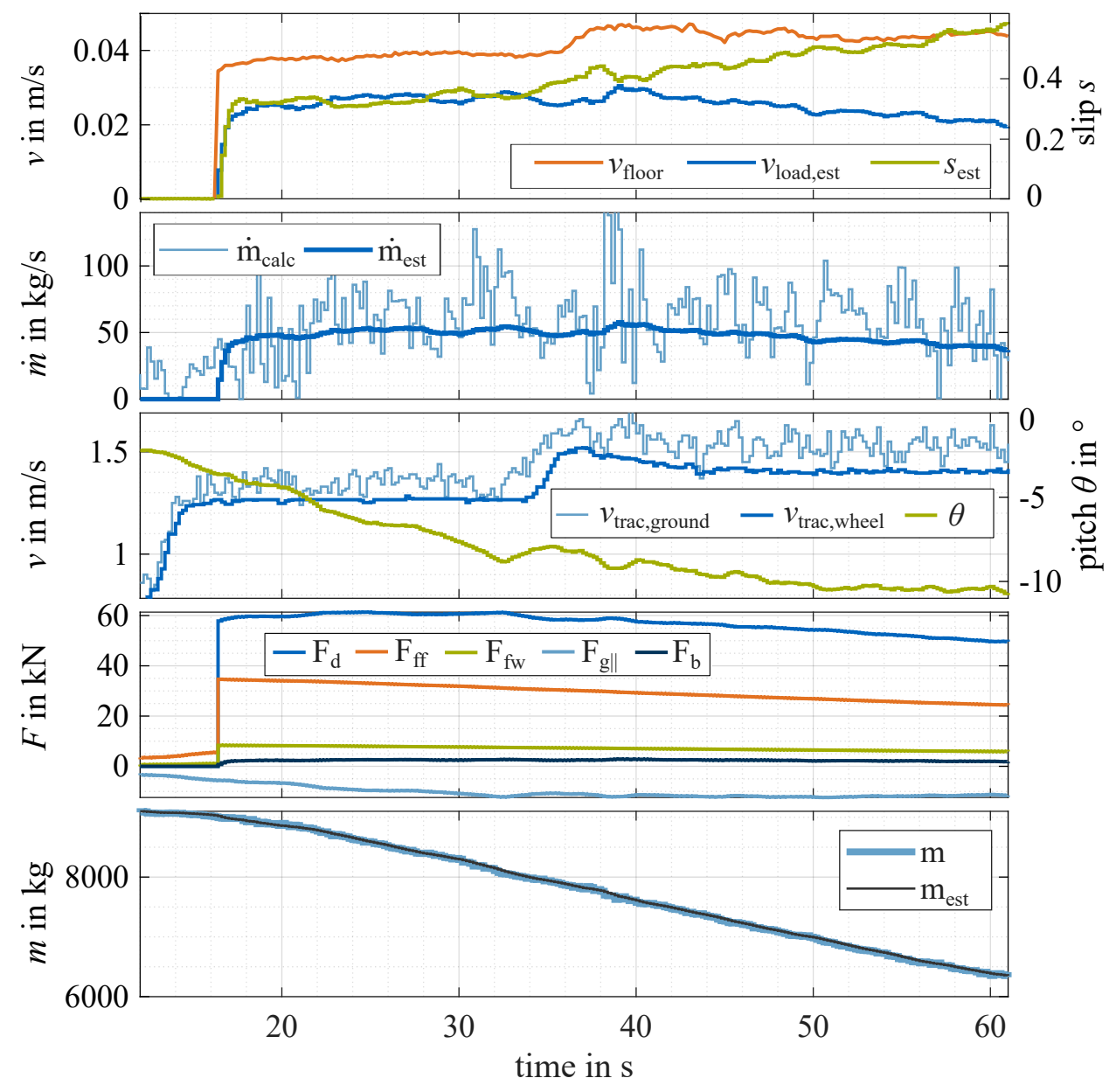

Fig. 7. EKF estimates and slip model on a downhill track; target application rate $16\,\mathrm{t}\cdot\mathrm{ha}^{-1}$.

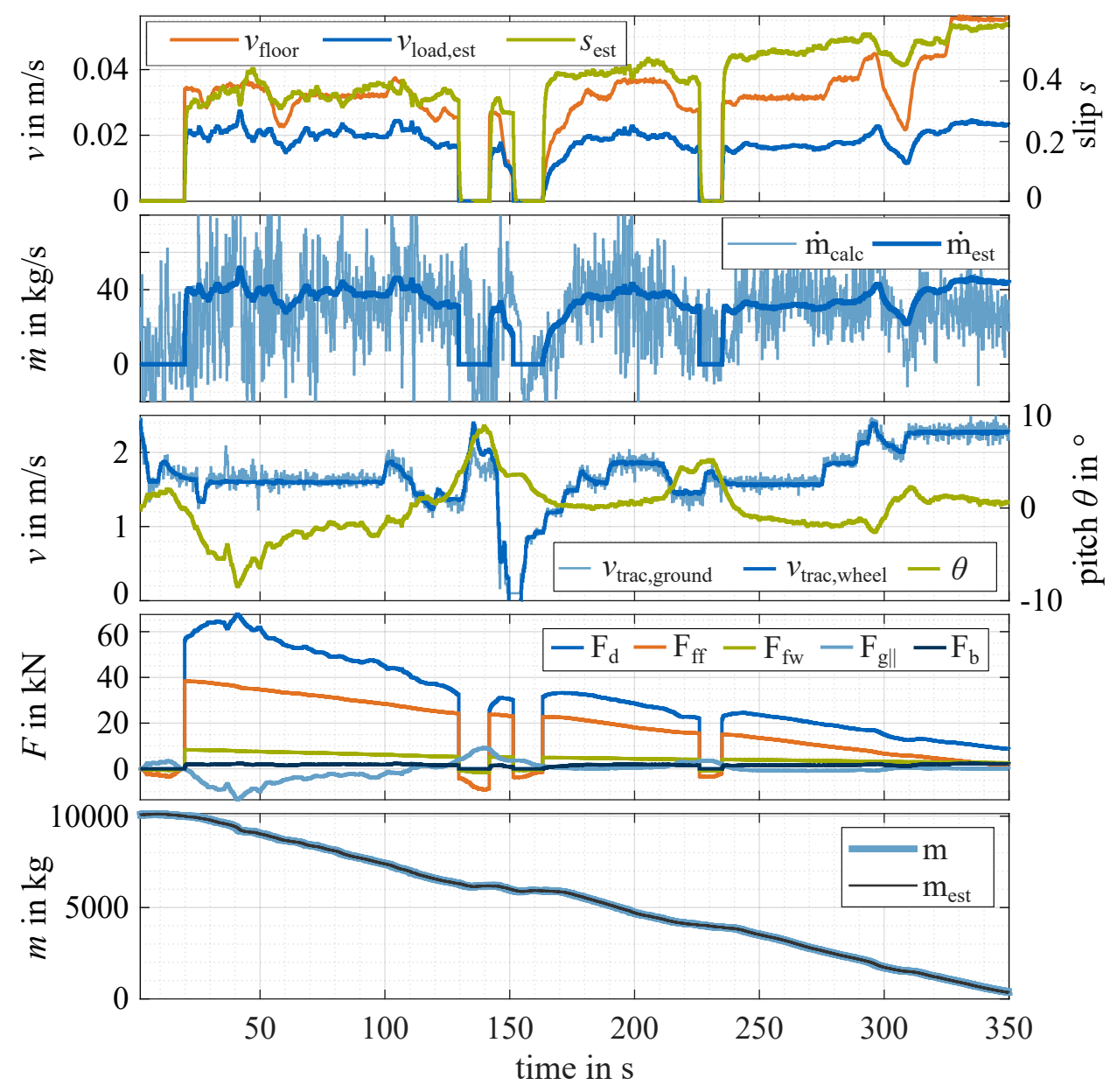

Fig. 8. Complete load cycle from full to empty; target application rate $10\,\mathrm{t}\cdot\mathrm{ha}^{-1}$.

## 5. DISCUSSION

The results demonstrate that the proposed algorithm robustly and accurately estimates both slip and mass flow under realistic operational conditions and undulating terrain, requiring only an IMU in addition to sensors commonly available on commercial spreaders. Compared to previous studies discussed in Sec. 1, mass flow can be estimated in real time without introducing significant filter dynamics. This enables integration into spreader transport floor speed control, facilitating more precise application rate control. Slope effects become relevant beyond 8°, which occurs regularly in agricultural terrain; the slight overestimation of the downslope force nonetheless warrants refinement.

Limitations remain: force-model ground truth is unavailable, so the model is only indirectly validated via the on-board weight and floor-speed sensors; material parameters were calibrated for a single manure type and may need adjustment for other products, though EKF measurement fusion limits parameter sensitivity; and start-up and near-empty phases, when the material does not fully contact the beaters, remain outside the estimator's scope.

## 6. CONCLUSION

An EKF-based estimator for slip and mass flow during solid-manure unloading was developed and validated on real-world data. The key contribution is the adaptation of the Bekker–Wong terramechanics formulation to conveyor–material interaction. The resulting real-time 10 Hz estimates enable integration into transport-floor speed control for improved variable-rate application.

## ACKNOWLEDGEMENTS

The authors thank Ludwig Bergmann GmbH (Goldenstedt, Germany) for providing the spreader and technical support.

## DECLARATION OF GENERATIVE AI

During the preparation of this work, the authors used Claude (Anthropic) in order to assist with light editing for spelling and grammar. After using these tools, the authors reviewed and edited the content as needed and take full responsibility for the content of the publication.